\documentclass{sig-alternate-05-2015}
\usepackage{tikz}
\usetikzlibrary{shapes,arrows,external}
\usepackage[hidelinks]{hyperref}
\usepackage{booktabs}
\usepackage{pifont}

\usepackage{scrextend}
\usepackage{array}
\usepackage{listings}
\definecolor{color:keyword}{rgb}{0.53,0.05,0.05}
\definecolor{color:comment}{rgb}{0.25,0.37,0.75}
\definecolor{color:string}{rgb}{0.87,0.0,0.0}
\usepackage{bold-extra}
\lstdefinelanguage{Jolie}{
morekeywords={
	provide,until,OneWay,RequestResponse,new,
	main,define,inputPort,outputPort,init,execution,include,
	cset,if,else,csets,interface,extender,forward,courier,with,type,throws,global,constants,for,
foreach,while,int,double,raw,void,undefined,string,long,bool,any,single,
sequential,concurrent,Jolie,Java,JavaScript,embedded,Location,Protocol,
Interfaces,Aggregates,scope,install,cH,comp,throw,this,default,synchronized,
nullProcess,false,true,Redirects
},
sensitive=true,
morecomment=[l]{//},
morecomment=[s]{/*}{*/},
morestring=[b]",
otherkeywords={;,|,:}
}

\newcommand{\smallpar}[1]{\smallskip \noindent \textbf{#1.}\ }

\usepackage{breakurl}
\begin{document}


\doi{}

\isbn{}



%

\title{
Circuit Breakers, Discovery, and API Gateways
in Microservices
}

%
%
%
%
%

\numberofauthors{2} 
%
\author{
\alignauthor Fabrizio Montesi\\
       \affaddr{Department of Mathematics and Computer Science}\\
       \affaddr{University of Southern Denmark}\\
       \affaddr{5230 Odense M, Denmark}\\
       \email{fmontesi@imada.sdu.dk}
\alignauthor Janine Weber\\
       \affaddr{Department of Mathematics and Computer Science}\\
       \affaddr{University of Southern Denmark}\\
       \affaddr{5230 Odense M, Denmark}\\
       \email{jaweb10@student.sdu.dk}
}


\maketitle
\begin{abstract}
We review some of the most widely used patterns for the programming of microservices: circuit breaker, service
discovery, and API gateway.
By systematically analysing different deployment strategies for these patterns, we reach new insight especially
for the application of circuit breakers.
We also evaluate the applicability of Jolie, a language for the programming of microservices, for these patterns and
report on other standard frameworks offering similar solutions. Finally, considerations for future developments are
presented.
\end{abstract}

%
%
%
%

%
%
\printccsdesc


\keywords{Design Patterns; Microservices; SOA}

\section{Introduction}

In the \emph{microservices} architectural style~\cite{FL14}, the components of an application are
autonomous services that execute independently and communicate via message
passing~\cite{DGLMMMS16}. This style is inspired by Service-Oriented Architecture (SOA). The key difference between the
two approaches lies in granularity. Even if services in SOA applications also communicate via message passing,
differently from microservices the internal components of each application are all part of a single executable artifact,
called a monolith. Consider, for example, a service in an SOA that includes an Auth(entication) and an Email
component.
In microservices, the two components would also be external services, each with
its own database.
Such services are sometimes called microservices, to point out that
they are designed using the microservices style.

Some key advantages that come from the granularity of a microservice architecture (MSA for short) are
(see~\cite{DGLMMMS16} for a more thorough analysis):
\begin{itemize}
\item Components can be deployed separate, allowing for the independent management of their respective lifecycles.
\item New versions of components can be gradually introduced in a system, by deploying them side to side with previous
versions. This advantage can be incorporated in Continuous Integration.
\item Components can be more specialised, since they can be written in different technologies -- as long as these
technologies support interaction with the other technologies used in the same MSA, via message passing.
\item Scaling a microservice architecture does not imply a duplication of
  all its components and developers can conveniently deploy/dispose instances
  of services with respect to their load~\cite{GGGMM16}.
\end{itemize}

Microservices has become increasingly popular over the last few years.
Some companies, such as Netflix and Amazon, have successfully become early adopters of microservices in the setting of 
large-scale
software systems.

Alas, the adoption of microservices also comes with its own set of issues. While many of these issues are inherited
directly from distributed systems, they are also exacerbated by the high degree of distribution of an MSA and the fact
that we must take them into account even for the composition of internal components. Some key issues include:
\begin{itemize}
\item Interactions among microservices happen via message passing, which introduces the possibilities of communication 
failures and timeouts among components.
\item Services may become overloaded, because of too many concurrent client requests or resources being kept busy
while waiting for replies from other services. This may easily trigger disastrous cascading
failures.
\item Microservices are optimised for cloud computing, so some services may be relocated at runtime.
\item Microservices can use different technologies, enabling specialisation to specific clients and tasks. Also,
MSAs are flexible and their APIs may change over time.
Therefore, MSAs should be supported by means for the rapid publishing of new APIs of different natures.
\end{itemize}

The solutions to these problems come from different sources. The first two issues can be solved with patterns from
highly-available systems, the third by using discovery and deployment mechanisms studied for SOA. Furthermore, many of
such solutions are aimed at specific applications and are provided by different vendors (e.g., the API gateways by
Amazon~\cite{j:AWSAPIG} and
Netflix~\cite{j:NOSSZ}).
Differently, our aim in this work is to discuss the principles behind the respective solutions for these
problems. Our main contribution is a homogeneous overview of a set of common solutions that microservices
developers should be aware of, equipped with novel insight related to their specific applications to microservices. We
also demonstrate how the solutions that carry novelty in the setting of microservices can be prototyped using
constructs developed for service composition in
Jolie~\cite{MGZ14}, a native microservice programming
language~\cite{jolie:website}. This is useful both as a reference
and as an evaluation of Jolie itself.

\smallpar{Structure of the paper}
We start our investigation from the pattern of circuit breaker, which deals
with the first two issues given above (\S~\ref{sec:cb}). While circuit breakers are typically employed client-side, we
observe that they can be useful also at other locations, and develop a Jolie prototype that can be transparently reused
regardless of where it is located.
We proceed by reporting on service discovery, which deals with the third issue, and discuss briefly two main
strategies: client- and server-based discovery (\S~\ref{sec:discovery}).
Our survey ends with a discussion of API Gateway, a pattern for the rapid deployment of new APIs in MSAs
(\S~\ref{sec:gateway}).
Related work is in \S~\ref{sec:related}. We report on conclusions and
future work in \S~\ref{sec:conclusions}.

\section{Circuit Breakers}\label{sec:cb}
%
%
Given enough incoming requests, even the most reliable of services will eventually exhaust its capabilities
and fail. Failure in an MSA is inevitable, and should be embraced with precaution rather than ignored.
What makes matters complex is that, in an MSA, a failing service probably has other services that depend on it. What
happens if our failing service becomes unresponsive? If we do not properly plan for this event, we risk all the other
services that rely on it to become unresponsive, too. This is called a cascading failure.

The circuit breaker pattern is aimed at preventing the failure of a single component to cascade beyond its boundaries,
and thereby bring the entire system down with it. The motto here is to fail fast: when a service becomes unresponsive,
its invokers should stop waiting for it, assume the worst, and start dealing with the fact that the failing service may
be unavailable.
Thus, circuit breakers contribute to the stability and resilience of both clients and services:
clients limit their waste of resources on trying to access unresponsive services, and overloaded services are given a
chance to recover by finishing some of the tasks they are currently processing.

\tikzset{cross/.style={cross out, draw=black,thick, minimum size=10*(#1-\pgflinewidth), inner sep=1pt, outer sep=1pt,
color=red, minimum height = 0.4cm},
	cross/.default={1pt}}
\begin{figure}
	\pgfdeclarelayer{background}
	\pgfdeclarelayer{foreground}
	\pgfsetlayers{background,main,foreground}
	\tikzstyle{state} = [draw, rounded corners,
	text centered, minimum height=3em, minimum width=4em]
	\centering
	\centering
	\begin{tikzpicture}[
	]
	\node (closed) [state, fill=green!20] {\textbf{closed}};
	\node (open) [state, fill=red!20, right of=closed, node distance =4.5cm] {\textbf{open}};
	\node (halfopen) [state, fill=yellow!20, below of=open, node distance =2.5cm] {\textbf{half-open}};
	\draw [->] (closed) -- (open) node[above,midway,align=center] {\textit{trip breaker}\\\textit{$[$threshold 
reached$]$}};
	\draw [->] (halfopen) -- (closed) node[left,midway] {\textit{success}};
	\draw [transform canvas={xshift=-0.3cm},->] (halfopen) -- (open) node[left,midway] {\textit{fail}};
	\draw [transform canvas={xshift=0.3cm},->] (open) -- (halfopen) node[right,midway,align=center] 
{\textit{attempt}\\\textit{reset}};
	\path[]	(open)   edge[loop, in=60,out=120, looseness=3,->] node[above]  {\textit{fail fast}} (open);
	\path[]	(closed)   edge[loop, in=60,out=120, looseness=3,<-] node[above]  {\textit{success}} (closed);
	\path[]	(closed)   edge[loop, in=240,out=300, looseness=3,<-] node[below]  {\textit{fail}} (closed);
	\end{tikzpicture}
	\caption{Circuit Breaker State Diagram.}
	\label{cb-fsm}
\end{figure}
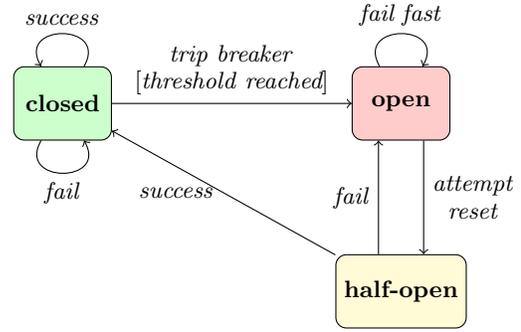
Concretely, a circuit breaker works by wrapping calls towards a target service and monitoring their failure rates.
The idea is that when the target service becomes too slow or replies too often with faults, the circuit breaker
will trip and future invocations from the client will immediately return a fault.
meeting
More specifically, the pattern can be implemented as a finite-state machine, depicted in Figure \ref{cb-fsm}. We
describe these states in the following.
\begin{description}
	\item[Closed:] Requests are passed to the target service. Faults caused by the requested
operation such
as exceptions or timeouts increase the circuit breaker's respective failure and timeout counters. Should these counters
exceed a specified threshold, or should another predefined criteria be met (e.g., a particular fault was
raised), the breaker is tripped and transitions into the open state.
	\item[Open:] Requests are not passed to the target service. Instead, a failure message is immediately given to the
client as reply. Potential fallback mechanisms can
be called to handle the failure. The circuit breaker can transition from the open to the half-open state, either by
periodically pinging the service to check for when it becomes responsive again, or after a specified amount of
time.
	\item[Half-Open:] While in this state, a limited number of requests are allowed through to the service. Provided
that the target service sends back successful replies, the circuit breaker is reset back into the
closed state, and its failure and timeout counters are reset. Should,
however, any of the requests fail while in the half-open state, the circuit breaker transitions back into the open
state.
\end{description}

The state transitions for circuit breakers are generally controlled by a set of parameters, which typically
includes those described in Table~\ref{t:cbp}.
%

\begin{table*}
	\centering
	\begin{tabular}{|c|c|c|} \hline
		\textbf{Parameter} & \textbf{Example Value} & \textbf{Explanation}\\
		\hline
		callTO & 20s & timeout the client request after 20 seconds without a response from the server\\
		rollingWindow & 60s & monitor errors over a rolling window of 60 seconds\\
		tripThreshold & 5\% & open the circuit if the error rate gets $\geq$ 5\%\\
		resetTO & 30s & attempt to reset the circuit after 30 seconds of opening the circuit\\
		\hline\end{tabular}
	\caption{Example of Circuit Breaker Parameters}
	\label{t:cbp}
\end{table*}

\paragraph{Deployment}
One of the most famous implementations of circuit breakers is provided by the Hystrix
library~\cite{j:NOSSH}, which allows to wrap
Java code in a procedure that will be controlled by a circuit breaker. The idea is that the circuit breaker is used
directly inside of the client.
Here, we make the (novel) observation that it makes sense to deploy a circuit breaker also in other places than just
inside of clients. Specifically, circuit breakers may also be introduced on the side of services, or in proxies that
operate between clients and services. Each strategy has its own advantages and disadvantages. Therefore, we envision
that practical applications should combine them. We start from the standard strategy of
client-side circuit breakers, for reference, and then move to the other ones.

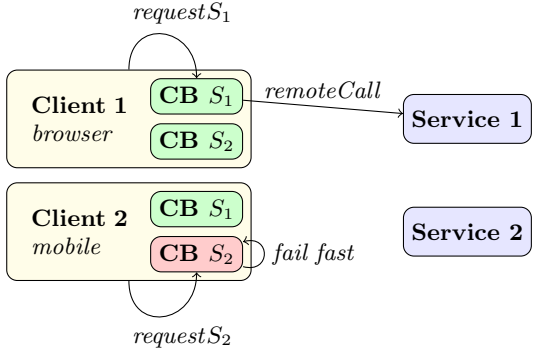
\begin{figure}
	\pgfdeclarelayer{background}
	\pgfdeclarelayer{foreground}
	\pgfsetlayers{background,main,foreground}
	\tikzstyle{client} = [draw, rounded corners,  left,
	minimum height=4em, minimum width=10em]
	\tikzstyle{service} = [draw, rounded corners,  left,
	minimum height=2em, minimum width=5em]
	\tikzstyle{cb} = [draw, rounded corners,  left]
	\centering
	\centering
	\begin{tikzpicture}[
	]
	\node (client1) [client, fill=yellow!10, text width = 8em] {\textbf{Client 1}\\\textit{browser}};
	\node (c1cb1) [cb, above of=client1, xshift=0.9cm,yshift=-0.7cm,fill=green!20,align=center] {\textbf{CB $S_1$}};
	\node (c1cb2) [cb, below of=client1, xshift=0.9cm,yshift=0.7cm,fill=green!20,align=center] {\textbf{CB $S_2$}};
	\node (client2) [client, below of=client1, node distance = 1.5cm, fill=yellow!10, text width = 8em] {\textbf{Client 
2}\\\textit{mobile}};
	\node (c2cb1) [cb, above of=client2, xshift=0.9cm,yshift=-0.7cm,fill=green!20,align=center] {\textbf{CB $S_1$}};
	\node (c2cb2) [cb, below of=client2, xshift=0.9cm,yshift=0.7cm,fill=red!20,align=center] {\textbf{CB $S_2$}};

	\node (service1) [service, right of = client1, node distance = 4.5cm, fill=blue!10] {\textbf{Service 1}};
	\node (service2) [service, right of = client2, node distance = 4.5cm, fill=blue!10] {\textbf{Service 2}};
	\draw [->] (c1cb1) -- (service1) node[above,midway,align=center] {\textit{remoteCall}};
	\path[]	(c2cb2)   edge[loop, in=15, out=345, looseness=3,->] (c2cb2) node[right,align=right,xshift=0.9cm] 
{\textit{fail fast}};
	\path[]	(client1)   edge[loop, in=90, out=90, looseness=2,->] (c1cb1) node[left,align=right, 
xshift=1.5cm,yshift=1.4cm] {\textit{request}$S_1$};
	\path[]	(client2)   edge[loop, in=270, out=270, looseness=2,->] (c2cb2) node[left,align=right, 
xshift=1.5cm,yshift=-1.4cm] {\textit{request}$S_2$};
	\end{tikzpicture}
	\caption{Client-side Circuit Breakers.}
	\label{cb-client}
\end{figure}

\smallpar{Client-side Circuit Breaker} The first deployment strategy is to place circuit breakers directly within
clients, as depicted in Figure~\ref{cb-client}. In this strategy, each client includes a separate circuit breaker for
intercepting calls to each external service that the client may call.
The strongest advantage of this strategy is that, when the circuit breaker is open, the target service will not receive
any messages from the client. This means that the service does not need to implement any similar protection mechanisms
of its own, relieving it from using resources for such mechanisms.
However, this requires two strong assumptions: we are able of forcing clients to use our circuit
breakers (e.g., access to the client source code); and, we are guaranteed that all clients are not malicious
(they will actually execute our code).
What if our scenario does not satisfy these requirements?
Another disadvantage is that the knowledge about the availability of a service is local to the client, depending on
how often requests are made to that service. For instance, in Figure~\ref{cb-client} \textit{Client 1} is unaware of
the unavailability status of \textit{Service 2}, whereas the circuit breaker of \textit{Client 2} for the same service
has been recently blown (denoted by the red colour). To counteract this issue, regular pings could be sent out to every
service in order to inquire about
their health status (but this functionality should then be supported by the services).

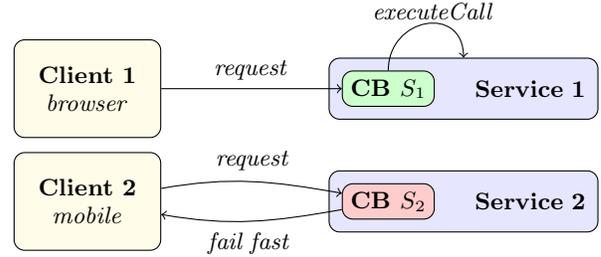
\begin{figure}
	\pgfdeclarelayer{background}
	\pgfdeclarelayer{foreground}
	\pgfsetlayers{background,main,foreground}
	\tikzstyle{proxy} = [draw, rounded corners,  left,
	minimum height=4em, minimum width=12em]
	\tikzstyle{client} = [draw, rounded corners,  left,
	minimum height=4em, minimum width=6em]
	\tikzstyle{service} = [draw, rounded corners,  left,
	minimum height=2.5em, minimum width=11em]
	\tikzstyle{cb} = [draw, rounded corners,  left]
	\centering
	\centering
	\begin{tikzpicture}[
	]
	\node (client1) [client, fill=yellow!10, align=center] {\textbf{Client 1}\\\textit{browser}};
	\node (client2) [client, below of =client1, node distance = 1.5cm, fill=yellow!10, align=center] {\textbf{Client 
2}\\\textit{mobile}};
	\node (service1) [service, right of = client1, node distance = 5cm, fill=blue!10, align=right, text width=10em] 
{\textbf{Service 1}};
	\node (service2) [service, right of = client2, node distance = 5cm, fill=blue!10, align=right, text width=10em] 
{\textbf{Service 2}};
	\node (s1cb) [cb, left of=service1,node distance = 1cm,fill=green!20,align=center] {\textbf{CB $S_1$}};
	\node (s2cb) [cb, left of=service2,node distance = 1cm,fill=red!20,align=center] {\textbf{CB $S_2$}};
	\draw [->] (client1) -- (s1cb) node[above,midway,align=center] {\textit{request}};
	\path[->]	(client2)   edge[loop, in=170, out=10, looseness=1] (s2cb) node[above,align=center,xshift=2.2cm, 
yshift=0.3cm] {\textit{request}};
	\path[->]	(s1cb)   edge[loop, in=90, out=90, looseness=1.8] (service1) node[above,align=center,xshift=0.6cm, 
yshift=0.8cm] {\textit{executeCall}};
	\path[->]	(s2cb)   edge[loop, out=190, in=350, looseness=1] (client2) node[below,align=center,xshift=-1.85cm, 
yshift=-0.3cm] {\textit{fail fast}};
	\end{tikzpicture}
	\caption{Service-side Circuit Breakers.}
	\label{cb-service}
\end{figure}

\smallpar{Service-side Circuit Breaker}
Circuit breakers can also be implemented on the side of services, as presented in Figure~\ref{cb-service}.
The idea is that all client invocations received by a service are first processed by an internal
circuit breaker, which decides whether the invocation should be processed or not.
A benefit here is that we do not make any of the assumptions necessary for client-side circuit breakers (in particular, 
clients can be
malicious). However, we have to change the behaviour of the service (e.g., by changing its source code); also,
the service uses resources to run the circuit breaker and receive messages even when the circuit breaker is open.
An interesting aspect here is that the service can see aggregate information about its responsiveness that encompasses
requests from all clients. A possible application is to develop a circuit breaker that throttles requests to
temporarily lighten the load on the service.

\begin{figure}
	\pgfdeclarelayer{background}
	\pgfdeclarelayer{foreground}
	\pgfsetlayers{background,main,foreground}
	\tikzstyle{proxy} = [draw, rounded corners,  left,
	minimum height=4em, minimum width=12em]
	\tikzstyle{client} = [draw, rounded corners,  left,
	minimum height=4em, minimum width=6em]
	\tikzstyle{service} = [draw, rounded corners,  left,
	minimum height=2em, minimum width=5em]
	\tikzstyle{cb} = [draw, rounded corners,  left]
	\centering
	\centering
	\begin{tikzpicture}[
	]
	\node (proxy) [proxy, fill=cyan!10, text width = 12em, align=center] {\textbf{Proxy}};
	\node (s1cb1) [cb, above of=proxy, xshift=1.3cm,yshift=-0.7cm,fill=green!20,align=center] {\textbf{CB $S_1$}};
	\node (s1cb2) [cb, below of=proxy, xshift=1.3cm,yshift=0.7cm,fill=red!20,align=center] {\textbf{CB $S_2$}};
	\node (c1cb1) [cb, above of=proxy, xshift=-1.3cm,yshift=-0.7cm,fill=green!20,align=center] {\textbf{CB $C_1$}};
	\node (c1cb2) [cb, below of=proxy, xshift=-1.3cm,yshift=0.7cm,fill=red!20,align=center] {\textbf{CB $C_2$}};
	\node (client1) [client, left of = proxy, yshift = 0.8cm, node distance = 4.5cm, fill=yellow!10, align=center] 
{\textbf{Client 1}\\\textit{browser}};
	\node (client2) [client, below of =client1, node distance = 1.5cm, fill=yellow!10, align=center] {\textbf{Client 
2}\\\textit{mobile}};
	\node (service1) [service, above left of = proxy, xshift=-0.5cm, yshift = 1.5cm, fill=blue!10] {\textbf{Service 1}};
	\node (service2) [service, above right of = proxy, xshift=0.5cm, yshift = 1.5cm, fill=blue!10] {\textbf{Service 2}};
	\draw [->] (client1) -- (proxy) node[above,midway,align=center] {\textit{request}\\$\{S_1,S_2\}$};
	\draw [->] (proxy) -- (service1) node[right,midway,align=center] {\textit{remoteCall}};
	\draw [->] (client2) -- (proxy) node[above,midway,align=center] {\textit{request}};

	\path[]	(c1cb2)   edge[loop, in=350, out=270, looseness=0,->] (client2) node[below,align=right,xshift=-1cm, 
yshift=-0.5cm] {\textit{fail fast}};
	\path[]	(s1cb2)   edge[loop, in=300, out=240, looseness=4,->] (s1cb2) node[below,align=right,yshift=-0.5cm] 
{\textit{fail fast}};
	\end{tikzpicture}
	\caption{Proxy with Circuit Breakers.}
	\label{cb-proxy}
\end{figure}
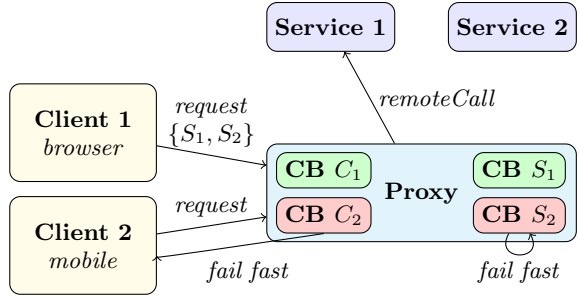

\smallpar{Proxy Circuit Breaker}
The last option that we present for the deployment of circuit breakers is a composition of the previous two
strategies. In this strategy, circuit breakers are deployed in a proxy service that sits between clients and
services, which handles all incoming and outgoing messages as displayed in Figure~\ref{cb-proxy}. The
proxy contains a circuit breaker for every client and every service within the system.\footnote{In systems where new
clients and/or services may join at runtime, circuit breakers may have to be created dynamically, but this is
orthogonal to our discussion.}
For any request from a client to a service to be allowed to go through, the respective circuit breakers of both client
and service must be closed. For instance, in the case illustrated in Figure~\ref{cb-proxy}, \textit{Client 1} is
allowed to send a request to \textit{Service 1}, but receives an exception when trying to call \textit{Service 2}.
Furthermore, \textit{Client 2} immediately gets denied any requests.
Observe that using a single proxy for multiple services introduces a network bottleneck, which in some cases plays well
with the system (e.g., in case the proxy can be deployed at a routing point) and other times it does not. In the latter
cases, it may be desirable to have one proxy for each target service. However, we will see that a proxy aggregating
multiple services makes sense in another pattern presented later in \S~\ref{sec:gateway}, the API Gateway.

Using proxies for deploying circuit breakers has two main benefits. First, this architectures simply requires
to configure clients to point to the proxy instead of the services directly. In many cases, this does not require
access to the client source code, but simply either a network reconfiguration or passing some location parameter to
clients to bind them correctly. It is also unnecessary to modify the code or configuration of the target services,
which can be seen as black boxes. Second, clients and services are equally protected from each other: clients are
made more resilient against faulty services, and services are shielded against cases in which a single client sends too
many requests. This also opens up the possibility of using shared knowledge among the circuit breakers, for
more refined strategies.

\paragraph{Implementation}
We sketch an implementation of a circuit breaker using Jolie, a native programming language for
microservices~\cite{MGZ14}.
The motivation for using Jolie is that all components in Jolie are (micro)services, and their definition is independent
of their deployment. As such, our prototype can be adopted in all deployment strategies that we reported in the
previous discussion, simply by loading it appropriately where desired. Specifically, this is because services in Jolie
can be deployed both as internal components that communicate using local memory or as distributed over a network. Our
definition is also independent of the interface of the target service, and the transport used for communicating messages
(we support all transports offered by Jolie, e.g., HTTP/JSON, HTTP/XML, and the binary protocol SODEP).

Our prototype, given in Figure~\ref{fig:cb_impl}, is simple enough that it can be discussed without assuming prior
knowledge of the Jolie language. We describe its functioning after its code, given in the following, and describe the
necessary Jolie concepts as we encounter them.

\lstset{literate={\\Loc}{{$Loc$}}2
{\\Proto}{{$Proto$}}3}
\lstset{frame=none}
\begin{figure*}[t]
\begin{minipage}[t]{.49\textwidth}
\begin{lstlisting}
outputPort TargetSrv { ... }

interface extender CBIfaceExt {
RequestResponse:
	*(void)(void) throws CBFault
}

inputPort CB {
Location: \Loc		Protocol: \Proto
Aggregates: TargetSrv with CBIfaceExt
}

define trip { state = Open; resetTO }

define checkErrorRate {
	if ( state == Closed ) {
		shouldTrip@Stats()( shouldTrip );
		if ( shouldTrip ) { trip }
	} else if ( state == HalfOpen ) {
		trip
	}
}

define forwardMsg {
	callTO;
	install( default =>
		cancelCallTO; failure@Stats();
		checkErrorRate );
	forward( request )( response );
	success@Stats(); cancelCallTO
}
\end{lstlisting}
\end{minipage}
\hfill
\begin{minipage}[t]{.49\textwidth}
\begin{lstlisting}[firstnumber=32]
courier CB {
	[ TargetIface( request )( response ) ] {
		if ( state == Closed ) {
			forwardMsg
		} else if ( state == Open ) {
			throw( CBFault )
		} else if ( state == HalfOpen ) {
			checkRate@Stats()( canPass );
			if ( canPass ) {
				forwardMsg
			} else {
				throw( CBFault )
			}
		}
	}
}

main {
	[ callTO() ] {
		timeout@Stats(); checkErrorRate
	}

	[ resetTO() ] {
		if ( state == Open ) {
			reset@Stats(); state = HalfOpen
		}
	}
}
\end{lstlisting}
\end{minipage}
\caption{Circuit Breaker Implementation in Jolie.}
\label{fig:cb_impl}
\end{figure*}

We describe the prototype. The underlying idea is simple: our program defines a service that intercepts all calls
from a client to a target service, applying the logic of a circuit breaker. The \lstinline+outputPort+
\lstinline+TargetSrv+ (Line 1) contains the binding information towards the target service of the circuit breaker. We
omit its definition, since this changes depending on the deployment of the circuit breaker. To deploy the circuit
breaker client-side or in a proxy, then \lstinline+TargetSrv+ would point to
a remote location. For the service-side case, instead, \lstinline+TargetSrv+ would point to a local memory location.
In Lines 8--11, the \lstinline+inputPort+ \lstinline+CB+ deploys the input endpoint that will receive
client messages.
We abstract from the location where it is concretely deployed and the transport protocol that it uses; these
are just configuration parameters, denoted $Loc$ and $Proto$.
The \lstinline+Aggregates+ part is key: it instructs Jolie that all client messages received by the circuit
breaker (on input port \lstinline+CB+) for an operation declared by the target service (\lstinline+TargetSrv+)
will be forwarded to the latter. The part \lstinline+with CBIfaceExt+ declares that the types of all operations are
augmented according to the definition of \lstinline+CBIfaceExt+, given in Lines 3--6. Specifically,
\lstinline+CBIfaceExt+ states that all operations can now also throw the fault \lstinline+CBFault+, which we will use
to notify clients of failures generated by the circuit breaker.
Jolie allows us to write arbitrary code to process the messages intercepted from clients to the target service, called
a \lstinline+courier+ behaviour.
We define this code in Lines 32--47. In Line 33, we state that we want to intercept all messages for an
operation defined in the interface of our target service (\lstinline+TargetIface+); \lstinline+request+ is the variable
that stores the client message, and \lstinline+response+ is the variable that will be used at the end to send the
response to the client. We then implement the circuit breaker state machine, using the (global)
variable \lstinline+state+ to store our state. We assume that there exists an internal \lstinline+Stats+ component that
stores and computes the statistics used in the logic of a circuit breaker, configured accordingly to the
tripThreshold and
rollingWindow parameters (from Table~\ref{t:cbp}). We report the different cases depending on the current
state.
\begin{description}
 \item[Closed (Line 35)] We call procedure \lstinline+forwardMsg+, defined in Lines 24--31. The procedure starts by
calling another procedure \lstinline+callTO+ (omitted here), which starts a timer with duration set by the callTO
parameter.
We then install a fault handler, which will be executed in case invoking the target services raises an error. In case
of error, the handler would cancel the call timer (Line 27), register the failure in \lstinline+Stats+, and check
whether we should change state by invoking procedure \lstinline+checkErrorRate+. The latter is a simple procedure
(Lines 15--22), which if we are currently in a closed state asks \lstinline+Stats+ whether we should trip the circuit
breaker, based on the data accumulated so far about successes, timeouts, and failures.
In Line 29, we \lstinline+forward+ the message from the client to the target service. If we are successful,
we register the success in \lstinline+Stats+ and cancel the call timer (Line 30).

What if the call timer expires before it is cancelled (either by a success or an error raised by the
\lstinline+forward+ statement)? In this case, a message for operation \lstinline+callTO+ will be sent to
our circuit breaker. This is handled in the \lstinline+main+ procedure of the service, Lines 49--59, where if we
receive a message for \lstinline+callTO+ we update the internal statistics by registering that a timeout occurred and
then check the error rate of the service (which may cause the circuit breaker to trip).

\item[Open (Line 37)]
While in this state, the circuit breaker does not forward client requests and instead replies directly with a message
containing fault \lstinline+CBFault+.

\item[Half-Open (Lines 39-44)]
In this state, we ask \lstinline+Stats+ whether the message can pass (Line 39). If so, we proceed with
\lstinline+forwardMsg+ as in the open state. Otherwise, we send back to the client a fault \lstinline+CBFault+ as in
the open state.

\end{description}

The procedure used to trip the circuit breaker is \lstinline+trip+ (Line 13), which also starts a reset timer.
When this timer expires, operation \lstinline+resetTO+ will be invoked and make the transition to the half-open
state (Line 56).

\section{Service Discovery}\label{sec:discovery}

In practice, the location of a microservice may not be statically known at design time.
This is because microservices may be deployed in a cloud-based system, which could replicate and relocate services at
runtime. Therefore, a participant in an MSA may need to employ a service discovery mechanism.
This is typically achieved with the same idea adopted for Service-Oriented Architecture (SOA), i.e., by using a
service registry. A service registry is a service that can be used by other components to retrieve binding information
about other components.
Microservices talk to the registry in order to publish their locations (or that of other services), whereas clients
address the registry to discover registered services.

In principle, there is no difference between using a service registry in an SOA or in an MSA. However, in practice,
service discovery for SOA is often implemented as part of an Enterprise Service Bus (ESB), which acts as a central
coordination point for service communications. Instead, in microservices, there are still no standards but rather
various custom implementations (e.g., Eureka~\cite{j:NOSSE} and AWS Elastic Load Balancing~\cite{j:AWSELB}).
Nevertheless, we can distinguish between two main implementation strategies: client-side and server-side discovery.

In the client-side discovery pattern, depicted in Figure~\ref{clientside-discovery},
the client is aware that services do not have fixed locations. It thus queries the service registry for the location of
all the services that it needs. Thereafter, the client contacts the target services directly.
This architecture is simple, but it requires that clients are designed to follow this methodology.
The implementation of clients thus becomes more complex, since it has to implement the discovery logic. This logic
needs to be replicated for each programming language and/or framework used for the implementation of clients.

In the alternative server-side discovery pattern, displayed in Figure~\ref{serverside-discovery}, we delegate the
discovery logic to a dedicated router service. The client exclusively talks to the router responsible for the services,
which is set at a fixed location. Upon receiving a request, the router talks to the service registry to discover the
requested service, and then forwards the client request to the latter.
Contrary to client-side discovery, this pattern does not require clients to be aware of the fluid deployment
of microservices. However, the programmer needs to deploy an additional service (the router) that will consume
resources.

Often, both patterns are present within large-scale applications.
A server-side discovery structure exposes the public services to the outside world, whereas the client-side discovery
pattern handles server- or cluster-internal interactions.
\begin{figure}[t]
	\pgfdeclarelayer{background}
	\pgfdeclarelayer{foreground}
	\pgfsetlayers{background,main,foreground}
	\tikzstyle{proxy} = [draw, rounded corners,  left,
	minimum height=4em, minimum width=6em]
	\tikzstyle{client} = [draw, rounded corners,  left,
	minimum height=5em, minimum width=10em]
	\tikzstyle{service} = [draw, rounded corners,  left,
	minimum height=2em, minimum width=5em]
	\tikzstyle{cb} = [draw, rounded corners,  left]
	\centering
	\centering
	\begin{tikzpicture}[
	]
	\node (client1) [client,fill=yellow!10, text width=8.5em,align=left] {\textbf{Service}\\\textbf{Client}};
	\node (client2) [cb, right of =client1, node distance = 0.7cm, fill=orange!10, align=center] 
{\textit{registry-}\\\textit{aware}\\\textit{Client}};
	\node (registry) [proxy, right of = client1, node distance = 3.3cm, fill=cyan!10, yshift = -2cm, align = center] 
{\textbf{Service}\\\textbf{Registry}};
	\node (service1) [service, right of = client1,  node distance = 6cm, fill=blue!10] {\textbf{Service 1}};
	\node (service2) [service, below of = service1, node distance = 1.7cm, fill=blue!10] {\textbf{Service 2}};
	\draw [->] (client1) -- (registry) node[left,midway,align=center] {\textit{getRegistry}};
	\draw [dashed,<-] (registry) -- (service1) node[right,midway,align=center] {\textit{register}};
	\draw [dashed,<-] (registry) -- (service2) node[right,midway,xshift=0cm,yshift=0.1cm,align=center] {};
	\draw [->] (client1) -- (service1) node[below,xshift=-0.7cm,yshift=-0.1cm,midway,align=center] 
{\textit{remoteCall}};
	\draw [->] (client1) -- (service2) node[above,midway,align=center] {};
	\end{tikzpicture}
	\caption{Client-side Service Discovery.}
	\label{clientside-discovery}
\end{figure}
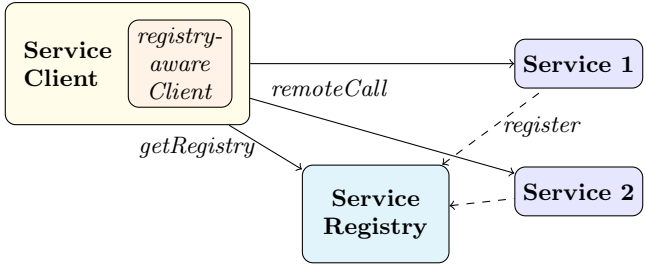
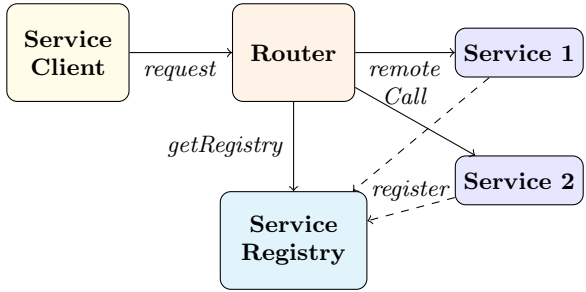
\begin{figure}[t]
	\pgfdeclarelayer{background}
	\pgfdeclarelayer{foreground}
	\pgfsetlayers{background,main,foreground}
	\tikzstyle{proxy} = [draw, rounded corners,  left,
	minimum height=4em, minimum width=6em]
	\tikzstyle{client} = [draw, rounded corners,  left,
	minimum height=4em, minimum width=5em]
	\tikzstyle{service} = [draw, rounded corners,  left,
	minimum height=2em, minimum width=5em]
	\tikzstyle{cb} = [draw, rounded corners,  left]
	\centering
	\centering
	\begin{tikzpicture}[
	]
	\node (client1) [client,fill=yellow!10,align=center] {\textbf{Service}\\\textbf{Client}};
	\node (router) [client, right of =client1, node distance = 3cm, fill=orange!10, align=center] {\textbf{Router}};
	\node (registry) [proxy, below of = router, node distance = 0.5cm, fill=cyan!10, yshift = -2cm, align = center] 
{\textbf{Service}\\\textbf{Registry}};
	\node (service1) [service, right of = router,  node distance = 3cm, fill=blue!10] {\textbf{Service 1}};
	\node (service2) [service, below of = service1, node distance = 1.7cm, fill=blue!10] {\textbf{Service 2}};
	\draw [->] (client1) -- (router) node[below,midway,align=center] {\textit{request}};
	\draw [->] (router) -- (registry) node[left,midway,align=center] {\textit{getRegistry}};
	\draw [dashed,->] (service1) -- (registry) node[above,midway,align=center] {};
	\draw [dashed,->] (service2) -- (registry) node[above,midway,yshift=0cm,align=center] {\textit{register}};
	\draw [->] (router) -- (service1) node[below,midway,align=center] {\textit{remote}\\\textit{Call}};
	\draw [->] (router) -- (service2) node[above,midway,align=center] {};
	\end{tikzpicture}
	\caption{Server-side Service Discovery.}
	\label{serverside-discovery}
\end{figure}
\section{API Gateways}\label{sec:gateway}
An MSA may need to serve different kinds of clients and user interfaces, such as
those found in web browsers and various smart devices (e.g., smartphones). Every client may have
different needs, depending on its target usage, form factor, and processing power. The needs of a client may even
change over time. For example, depending on the quality of its current network connection, a device may want to use
an API that is more or less network intensive -- for example, the description of a product may include more and
higher-quality resources, like pictures or embedded instructions.

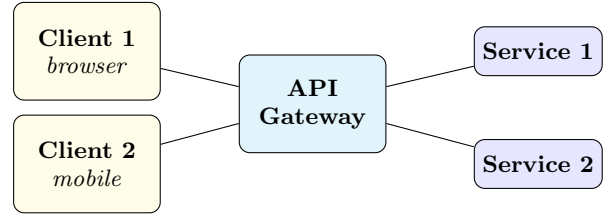
\begin{figure}[t]
	\pgfdeclarelayer{background}
	\pgfdeclarelayer{foreground}
	\pgfsetlayers{background,main,foreground}
	\tikzstyle{proxy} = [draw, rounded corners,  left,
	minimum height=4em, minimum width=6em]
	\tikzstyle{client} = [draw, rounded corners,  left,
	minimum height=4em, minimum width=6em]
	\tikzstyle{service} = [draw, rounded corners,  left,
	minimum height=2em, minimum width=5em]
	\tikzstyle{cb} = [draw, rounded corners,  left]
	\centering
	\centering
	\begin{tikzpicture}[
	]
	\node (client1) [client, left of = proxy, yshift = 0.8cm, node distance = 4.5cm, fill=yellow!10, align=center]
{\textbf{Client 1}\\\textit{browser}};
	\node (client2) [client, below of =client1, node distance = 1.5cm, fill=yellow!10, align=center] {\textbf{Client
2}\\\textit{mobile}};
	\node (service1) [service, right of = client1, node distance = 6cm, fill=blue!10] {\textbf{Service 1}};
	\node (service2) [service, right of = client2, node distance = 6cm, fill=blue!10] {\textbf{Service 2}};
	\node (api) [proxy, right of = client1, node distance = 3cm, fill=cyan!10, yshift = -0.7cm, align = center]
{\textbf{API}\\\textbf{Gateway}};
	\draw [] (client1) -- (api) node[above,midway,align=center] {};
	\draw [] (client2) -- (api) node[above,midway,align=center] {};
	\draw [] (service1) -- (api) node[above,midway,align=center] {};
	\draw [] (service2) -- (api) node[above,midway,align=center] {};
	\end{tikzpicture}
	\caption{API Gateway.}
	\label{api-gateway}
\end{figure}

The API Gateway is a service that addresses the issue of having clients of different natures. It is 
a single entry point that provides access to many APIs (Figure~\ref{api-gateway}).
An API Gateway provides functionalities for: publishing multiple APIs, each one dedicated to a different set of
clients; and, updating the set of published APIs at runtime (since developers may deploy new services during the
lifecycle of the MSA).

Since an API Gateway is an entry point for the MSA, it is natural to equip it with, e.g.,
service discovery, load balancing, monitoring, and security. Its position in the system is also ideal for adopting
the proxy circuit breaker pattern, by equipping the API Gateway with circuit breakers for clients and/or services.
The way in which these additional features are implemented depends on the implementation technology. In Jolie, this
would simply require to compose the specific services developed for the respective functionalities (e.g., our
circuit breaker prototype, or a service registry). Other technologies may need to add these features
directly inside of the codebase that implements the API Gateway.
Observe that for service discovery, both the client-side and server-side discovery patterns make sense here so we
should refer to their own advantages and disadvantages to choose between them. For client-side discovery, the API
Gateway should provide clients with access to the service registry -- while this access would still be brokered by the
gateway, it would still be the clients that select the services that they want (recall
Figure~\ref{clientside-discovery}).
For server-side discovery, the API Gateway would simply act as the router
(recall Figure~\ref{serverside-discovery}).

\begin{figure}[t]
\begin{lstlisting}
inputPort APIGateway {
Location: "socket://gateway.com:80/"
Protocol: http
Redirects:
	MobileAPI => MobileService,
	DesktopAPI => DesktopService
}

main {
	[ deploy( request )() {
		loadEmbeddedService@Runtime
													( ... )( ... );
		setRedirection@Runtime( ... )( ... )
	} ]
}
\end{lstlisting}
\caption{A sketch of an API Gateway in Jolie.}
\label{api-gateway-jolie}
\end{figure}

In Jolie, we can implement an API Gateway by using redirections~\cite{MGZ14}. A redirection makes an API available
under a specific name at a service input port. We sketch a prototype in Figure~\ref{api-gateway-jolie}.
In Lines 1--7, we declare the input port of the gateway (the location and protocol are just examples). The
input port provides two APIs, which can be reached respectively at the URLs
\burl{http://gateway.com:80/MobileAPI} and \burl{http://gateway.com:80/DesktopAPI}.
The service offers a \lstinline+deploy+ operation for deploying new APIs at runtime (Lines 10--14). We omit
the concrete data used in the operation. The idea is that, upon request, the gateway will use the Jolie standard
library to embed (run an internal service) all the necessary services for guarding the new API
(e.g., circuit breakers) and set a new redirection to publish the API.
Redirection in Jolie takes care of doing the necessary transformations between different communication transports, but 
sometimes
developers need to use special adapters to translate calls using ad-hoc procedures.
An advantage in this task is the possibility to use procedures written in different languages.
Jolie supports Jolie itself, JavaScript, and Java. For instance, if the mobile API in our redirection example
required an adapter written in JavaScript, we could use the following embedding instruction.
\begin{lstlisting}
embedded {
JavaScript:
	"mobile_adapter.js" in MobileService
}
\end{lstlisting}

\section{Related Work}\label{sec:related}
Being essentially distributed, microservices is founded on the well-known mechanism of message
passing. However, MSAs are much more involved than other distributed applications where services are implemented as
monoliths, because all internal components are subjects to potential communication failures and overloads.
So far, most proposals for dealing with these problems have been produced by practitioners, and therefore plenty of
valuable information has to be found in books and web resources.
We give here an overview of such solutions and compare to our work where appropriate.
A summary of the technologies that we mention is given in Table~\ref{t:sota}. We make an abuse of notation by reporting
Jolie along with frameworks, even though Jolie is actually a programming language and thus patterns must be
implemented. The idea is to point out that all such patterns are naturally supported by Jolie via its language
constructs.

\begin{table*}
	\centering
	\begin{tabular}{|>{\centering\arraybackslash}p{6cm}|>{\centering\arraybackslash}p{9.3cm}|} \hline
		\textbf{Pattern} & \textbf{Frameworks, Libraries \& Languages}\\
		\hline
		API Gateway & AWS\cite{j:AWSAPIG}, Netflix\cite{j:NOSSZ}, Nginx\cite{j:NGINXAPIG},
Jolie\cite{MGZ14,jolie:website} 	\\
		\hline
		Circuit Breaker & Hystrix\cite{j:NOSSH}, Akka\cite{j:AKKACB}, Jolie\cite{MGZ14,jolie:website} \\\hline
		Load Balancing \& Service Discovery & Nginx\cite{j:NGINXLB}, Ribbon\cite{j:NOSSR}, ELB\cite{j:AWSELB}, 
Eureka\cite{j:NOSSE}, etcd\cite{j:ETCD}, Zookeeper\cite{j:AZOO}, Marathon\cite{j:MARA}, Consul\cite{j:CON}, 
Jolie\cite{jolie:website} \\
		\hline
		Monitoring \& Metrics & Docker\cite{j:DOCK}, Hystrix\cite{j:NOSSH}, Lightbend\cite{j:LBM},
Marathon\cite{j:MARA}, Jolie\cite{MGZ14,jolie:website} \\
		\hline\end{tabular}
	\caption{State of the Art Microservice Frameworks, Libraries \& Languages.}
	\label{t:sota}
\end{table*}

\smallpar{Circuit Breakers}
Circuit breakers have first been popularised in~\cite{N07}, where their role is discussed in the context of
availability (resilience) for enterprise systems.

Akka~\cite{j:AKKACB} provides a circuit breaker implementation that supports basic configuration parameters, such as
call timeout, failure threshold and reset threshold.
Hystrix~\cite{j:NOSSH} is much more flexible and is currently one of the reference solutions: it supports rolling
statistics, fallback mechanisms, resource control, and control over the states and transitions of circuit
breakers.
Our circuit breaker prototype in Jolie (Figure~\ref{fig:cb_impl}) is of course not as mature
as these implementations, but it is interesting because it can be freely deployed in any of scenarios described in
\S~\ref{sec:cb}. Furthermore, our circuit breaker is parametric on the interface of the target service. This means that
if such interface changes over time (as can often happen in microservicese, e.g., by adding an operation), then the
circuit breaker can be re-adopted immediately without any changes.
Hystrix does not support this capability: supporting a new operation requires writing an additional implementation of
a \lstinline+HystrixCommand+. Importing this feature could be an interesting future development.

\smallpar{Service Discovery and Load Balancing}
Eureka~\cite{j:NOSSE} and Ribbon~\cite{j:NOSSR} combined together provide client-side load balancing and service
discovery. Amazon Web Services Elastic Load Balancing (AWS ELB), instead, implements balancing and discovery using
a server-side solution. Therefore, ELB is generally used to expose edge services to the public, and Eureka to handle
internal service communications. However, this also means that ELB can become a bottleneck. Eureka is more resilient,
as all information is cached by clients.
Discovery and load balancing can be achieved in Jolie using the techniques described in~\cite{DGGMM12,MGZ14}.

\smallpar{API Gateways}
Both Zuul~\cite{j:NOSSZ} and Amazon Web Services (AWS)~\cite{j:AWSAPIG} provide roughly the same
functionalities in their implementation (e.g., security, authentication, monitoring, and load balancing).
Zuul consists of different libraries (such as Eureka\cite{j:NOSSE},
Hystrix\cite{j:NOSSH}, and Ribbon\cite{j:NOSSR}), whereas AWS provides a single framework that may be quicker to
use in the beginning. However, since Zuul and all its dependencies are open source, it has the advantage in
customisation potential.
Redirections in Jolie as we used in \S~\ref{sec:gateway} do not take care of extra features such as security and
monitoring by themselves. The idea is that an API Gateway in Jolie should be composed with other patterns, e.g.,
circuit breakers, to achieve this extra features, keeping concerns separate. This is easy to do since all components
in Jolie are relocatable services that must define interfaces
for enabling composition.

\section{Conclusions and Future Work}\label{sec:conclusions}

We reviewed three mainstream mechanisms found in Microservice Architectures (MSAs): Circuit Breaker,
Service Discovery, and API Gateway.
We discussed different strategies for their implementation, and elicited the interplay between deployment
topologies and circuit breakers.

These patterns are emerging as essential for the reliability, ease of access, and flexibility of MSAs. Since
microservices is in its early development, we can expect more patterns like these to appear in the future.
It is interesting that these patterns are structural, in the sense that they
do not change the operations that services provided by developers offer, which are more custom to
the specific MSA at hand.
Being of this nature, their implementations benefit from parametricity to achieve reusability, as we have shown
for circuit breakers by using interface parametricity in Jolie.
However, their adoption also makes MSAs more complicated, and they influence the communication structures that will be
enacted in a system.
This suggests that methods for the programming and verification of communications among services should keep patterns
such as these into account.

Development methodologies for service communications typically employ choreographies for the description of service
protocols. Choreographies do not require central control, a critical feature for the scalability of MSAs.
Formal methods and languages based on choreographies have been developed for various purposes, including: documenting
systems using choreographies~\cite{WS-CDL,BPMN}; synthesising service implementations starting
from choreographies~\cite{M13:phd,CM13}; and, verifying safety properties of choreographies (e.g.,
deadlock-freedom)\cite{YHNN13,HYC16}.
The patterns that we considered cannot be readily implemented in these models. They require extensions to deal with
some necessary features, specifically: circuit breakers require timeouts, faults, and interface parametricity;
service discovery requires dynamic binding (the capability of connecting to a remote service whose location is
discovered at runtime); and, API gateways require the capability of loading new services at runtime.
There are promising works that deal with timeouts~\cite{BLY15}, faults~\cite{C09,CGY16}, dynamic binding~\cite{MY13},
and parametric behaviour~\cite{TST14,CLMSW16}.
However, all these works are not integrated with one another and a coherent choreography language that can capture, for
example, circuit breakers still has to appear. Therefore, designing a model capable of capturing the patterns that we
described represents interesting future work. Useful inspiration may be gained also from related work in the
area of process calculi and components, given their vicinity in the adopted techniques,
including~\cite{LZ05,GLMZ09,MS10,DGL13}.

\bibliography{biblio}

\end{document}